\documentclass[10pt,conference]{IEEEtran}
\usepackage{cite}
\usepackage{amsmath,amssymb,amsfonts}
\usepackage{algorithmic}
\usepackage{graphicx}
\usepackage{textcomp}
\usepackage{xcolor}
\usepackage[hyphens]{url}
\usepackage[hyphenbreaks]{breakurl}
\usepackage{booktabs}
\usepackage{pifont}
\usepackage{siunitx}
\usepackage{amsmath}
\usepackage{algorithm2e}
\usepackage{multirow}
\usepackage{fancyhdr}
\usepackage{float}
\usepackage{setspace}
\usepackage{listings}
\usepackage{balance}
\usepackage{wrapfig}
\usepackage{marginnote}
\usepackage{caption}
\usepackage{floatflt}
\definecolor{gfored}{rgb}{0.580, 0.050, 0.211}
\definecolor{ao}{rgb}{0.007, 0.520, 0.867}
\definecolor{yt}{rgb}{0.875, 0.568, 1.000}
\definecolor{moegi}{rgb}{0.357, 0.537, 0.188}
\definecolor{jl}{rgb}{1.0, 0.2, 0.8}
\definecolor{brown(web)}{rgb}{0.65, 0.16, 0.16}
\definecolor{bisque}{rgb}{1.0, 0.89, 0.77}

\newif\ifsqueezefigs
\squeezefigstrue

\ifsqueezefigs
\makeatletter
\g@addto@macro{\normalsize}{%
  \setlength{\abovedisplayskip}{2pt plus 1pt minus 1pt}
  \setlength{\belowdisplayskip}{2pt plus 1pt minus 1pt}
  \setlength{\abovedisplayshortskip}{0pt}
  \setlength{\belowdisplayshortskip}{0pt}
  \setlength{\intextsep}{2pt plus 1pt minus 1pt}
  \setlength{\textfloatsep}{3pt plus 1pt minus 1pt}
  \setlength{\dbltextfloatsep}{3pt plus 1pt minus 1pt}
  \setlength{\skip\footins}{4pt plus 1pt minus 1pt}}
\makeatother
\fi

\newif\ifdraft
\draftfalse

\ifdraft
    \usepackage[colorinlistoftodos,prependcaption,textsize=small]{todonotes}
    \paperwidth=\dimexpr \paperwidth + 4cm\relax
    \oddsidemargin=\dimexpr\oddsidemargin + 2cm\relax
    \evensidemargin=\dimexpr\evensidemargin + 2cm\relax
    \marginparwidth=\dimexpr \marginparwidth + 2cm\relax

    \newcommand{\ominline}[1]{\textcolor{red}{\textbf{[@om: }#1\textbf{]}}}
    \newcommand{\ombox}[1]{\todo[size=\scriptsize, linecolor=red, bordercolor=red, backgroundcolor=white]{\textcolor{red}{\textbf{@om:} #1}}}

    \newcommand{\agycomment}[1]{\todo[size=\scriptsize, linecolor=orange, bordercolor=orange, backgroundcolor=white]{\textcolor{gfored}{\textbf{@gy:} #1}}}
    \newcommand{\agyinline}[1]{\textcolor{gfored}{\textbf{[@agy: }#1\textbf{]}}}

    \newcommand{\atbcomment}[1]{\todo[size=\scriptsize, linecolor=orange, bordercolor=orange, backgroundcolor=white]{\textcolor{ao}{\textbf{@atb:} #1}}}

    \newcommand{\hluoinline}[1]{\textcolor{moegi}{\textbf{[@hluo: }#1\textbf{]}}}
    \newcommand{\hluobox}[1]{\todo[size=\scriptsize, linecolor=orange, bordercolor=orange, backgroundcolor=white]{\textcolor{moegi}{\textbf{@hluo:} #1}}}

    \newcommand{\yctcomment}[1]{\todo[size=\scriptsize, linecolor=orange, bordercolor=orange, backgroundcolor=white]{\textcolor{yt}{\textbf{@yct:} #1}}}

    \newcommand{\joel}[1]{\textcolor{jl}{#1}}
    \newcommand{\joelcomment}[1]{\todo[size=\scriptsize,linecolor=orange,bordercolor=orange,backgroundcolor=white]{\textcolor{jl}{\textbf{@joel:} #1}}}

\else

    \newcommand{\ominline}[1]{}
    \newcommand{\ombox}[1]{}

    \newcommand{\agycomment}[1]{}
    \newcommand{\agyinline}[1]{}

    \newcommand{\atbcomment}[1]{}

    \newcommand{\hluoinline}[1]{}
    \newcommand{\hluobox}[1]{}

    \newcommand{\yctcomment}[1]{}

    \newcommand{\joel}[1]{{#1}}
    \newcommand{\joelcomment}[1]{}

\fi

\newif\ifrebuttal
\rebuttaltrue

\ifrebuttal
\usepackage[colorinlistoftodos,prependcaption,textsize=small]{todonotes}

\definecolor{darkred}{rgb}{0.9, 0.0, 0.0}

\definecolor{darkblue}{rgb}{0.0, 0.0, 0.85}

\fi

\newcommand*\DRAMCMD[1]{\texttt{#1}}
\newcommand*\DRAMTIMING[1]{t\textsubscript{#1}}

\newcounter{obs}
\newcounter{tkw}
\usepackage{glossaries}

\newacronym{vdd}{$V_{DD}$}{supply voltage}
\newacronym{vpp}{$V_{PP}$}{wordline voltage}
\newacronym{vwl}{$V_{PP}$}{wordline voltage}
\newacronym{vgs}{$V_{GS}$}{gate-to-source voltage}
\newacronym{vth}{$V_{TH}$}{the voltage threshold that the bitline voltage should exceed for the activation to be reliably completed}
\newacronym{gnd}{$GND$}{ground}

\newacronym{ber}{$BER$}{the fraction of DRAM cells that experience bitflips in a DRAM row}
\newacronym{acmin}{$AC_{min}$}{the minimum number of total aggressor row activations to cause at least one bitflip}
\newacronym{ac}{$AC$}{activation count}
\newacronym{rblast}{$r_{Blast}$}{blast radius}
\newacronym{iqr}{$IQR$}{interquartile range}

\newacronym{trcd}{\DRAMTIMING{RCD}}{
{the minimum time between opening a row with an \DRAMCMD{ACT} command and accessing the row buffer}
}
\newacronym{trp}{\DRAMTIMING{RP}}{
{the minimum time between sending a \DRAMCMD{PRE} command and opening a row with an \DRAMCMD{ACT} command}
}
\newacronym{tras}{\DRAMTIMING{RAS}}{
{the minimum time between opening a row with an \DRAMCMD{ACT} command and closing the row with a \DRAMCMD{PRE} command}
}
\newacronym{trefi}{\DRAMTIMING{REFI}}{the \joel{default} time interval \joel{between consecutive \DRAMCMD{REF} commands}}
\newacronym{trefw}{\DRAMTIMING{REFW}}{the maximum time window between two consecutive refresh operations targeting {the same} row}

\def\BibTeX{{\rm B\kern-.05em{\sc i\kern-.025em b}\kern-.08em
    T\kern-.1667em\lower.7ex\hbox{E}\kern-.125emX}}

\author{
{Onur Mutlu}%
\vspace{-3pt}
\\
\emph{ETH Z{\"u}rich}%
\vspace{-20pt}
}

\title{\LARGE{\emph{Retrospective:} Flipping Bits in Memory Without Accessing Them: \\\vspace{-3pt} An Experimental Study of DRAM Disturbance Errors\vspace{-10pt}}}

\begin{document}
\maketitle
\thispagestyle{plain}
\pagestyle{plain}
\setstretch{0.765}
\begin{abstract}

Our ISCA 2014 paper~\cite{kim2014flipping} provided the first scientific and detailed characterization, analysis, and real-system demonstration of what is now popularly known as the RowHammer phenomenon (or vulnerability) in modern commodity DRAM chips, which are used as main memory in almost all modern computing systems. It experimentally demonstrated that more than 80\% of all DRAM modules we tested from the three major DRAM vendors were vulnerable to the RowHammer read disturbance phenomenon: one can predictably induce bitflips (i.e., data corruption) in real DRAM modules by repeatedly accessing a DRAM row and thus causing electrical disturbance to physically nearby rows. We showed that a simple unprivileged user-level program induced RowHammer bitflips in multiple real systems and suggested that a security attack can be built using this proof-of-concept to hijack control of the system or cause other harm. To solve the RowHammer problem, our paper examined seven different approaches (including a novel probabilistic approach that has very low cost), some of which influenced or were adopted in different industrial products. 

Many later works from various research communities examined RowHammer, building real security attacks, proposing new defenses, further analyzing the problem  at various (e.g., device/circuit, architecture, and system) levels, and exploiting RowHammer for various purposes (e.g., to reverse-engineer DRAM chips). Industry has worked to mitigate the problem, changing both memory controllers and DRAM standards/chips. Two major DRAM vendors finally wrote papers on the topic in 2023, describing their current approaches to mitigate RowHammer. Research \& development on RowHammer in both academia \& industry continues to be very active and fascinating. 

This short retrospective provides a brief analysis of our ISCA 2014 paper and its impact. We describe the circumstances that led to our paper, mention its influence on later works and products, describe the mindset change we believe it has helped enable in hardware security, and discuss our predictions for future.

\end{abstract}

\section{Background and Circumstances} 
\vspace{-4pt}

Our stumbling on the RowHammer problem and creation of its first scientific analysis happened as a result of a confluence of multiple factors. First, my group was working on DRAM technology scaling issues since late 2010. We were very interested in failure mechanisms that appear or worsen due to aggressive technology scaling. To study such issues (e.g., data retention errors~\cite{dram-isca2013}), we built an FPGA-based DRAM testing infrastructure~\cite{dram-isca2013} between 2011-2012, which we later open sourced as SoftMC~\cite{hassan2017softmc, softmc-safarigithub} and DRAM Bender~\cite{olgun2023drambender, safari-drambender}. Second, around the same timeframe, we were investigating similar technology scaling issues in flash memory using real NAND flash chips~\cite{cai2012error, cai-itj2013}. We knew read disturbance errors were significant in NAND flash memory~\cite{cai2012error, cai-iccd13, cai-itj2013, cai-dsn15, cai2017flashtbd} and were very interested in how prevalent they were in DRAM. Third, we were collaborating with Intel (e.g.,~\cite{dram-isca2013}) to understand and solve DRAM technology scaling problems and build our DRAM infrastructure. Three of my students and I spent the summer of 2012 at Intel to work closely with our collaborators (two are co-authors): during this time, we finalized the calibration and stabilization of our infrastructure and had significant technical discussions and experimentation on DRAM scaling problems.

Although there was awareness of the RowHammer problem in industry in 2012 (see Footnote~1 in~\cite{kim2014flipping}), there was no comprehensive experimental analysis and detailed real-system demonstration of it. We believed it was critical to provide a rigorous scientific analysis using a wide variety of DRAM chips and scientifically establish major characteristics and prevalence of RowHammer. Hence, in the summer of 2012, we set out to use our DRAM testing infrastructure to analyze RowHammer. Our initial results showed how widespread the read disturbance problem was across the (at the time) recent DRAM chips we tested, so we studied the problem comprehensively and developed many solutions to it. The resulting paper was submitted to MICRO in May 2013 but was rejected. We strengthened the results, especially of the mitigation mechanisms and the number of tested chips, and made the analysis more comprehensive before it was accepted to ISCA 2014 (2 of the 6 reviewers still rejected it for interesting reasons).

\vspace{-2pt}

\section{Major Contribution and Influence}
\vspace{-4pt}

The major contribution of our paper is the exposure and detailed analysis of a fundamental hardware failure mechanism that breaks memory isolation in real systems and thus has huge implications on system reliability, security, and safety. Our paper is a comprehensive study of a major DRAM technology scaling problem, RowHammer, including its first scientific analysis, experimental characterization, real system demonstration, and solutions with their evaluation. To our knowledge, RowHammer is the first example of a hardware failure mechanism that creates a significant and widespread system security vulnerability~\cite{onur-date17, mutlu2019retrospective, mutlu2023fundamentally, dullien-keynote-ccdcoe-2018}, as our ISCA 2014 paper suggested. 

Our work has had large influence on both industry \& academia. Individual follow-on works are many to list here; we refer the reader to longer invited retrospectives we wrote~\cite{onur-date17, mutlu2019retrospective, mutlu2023fundamentally}. We give major examples of influence, focusing on RowHammer's effect on the collective mindset of security research and major industry milestones related to RowHammer. 

{\em RowHammer Attacks \& Mindset Shift in Hardware Security.} Our demonstration that one can easily and predictably induce bitflips in commodity DRAM chips using a real user-level program enabled a major mindset shift in hardware security. It showed that general-purpose hardware is fallible in a very widespread manner and its problems are exploitable. Tens of works (see~\cite{mutlu2023fundamentally, mutlu2019retrospective}) built directly on our work to exploit RowHammer bitflips to develop many attacks that compromise system integrity and confidentiality, starting from the first RowHammer exploit by Google Project Zero in 2015~\cite{google-project-zero, seaborn2015exploiting} to recent works in 2022-2023 (e.g.,~\cite{rakin2022deepsteal, jolt23recovering}). These attacks showed increasingly sophisticated ways by which an unprivileged attacker can exploit RowHammer bitflips to circumvent memory protection and gain complete control of a system (e.g.,~\cite{google-project-zero, gruss2016rowhammer, van2016drammer, xiao2016one, flip-feng-shui, tatar2018throwhammer, lipp2018nethammer, cojocar2019eccploit, deridder2021smash, jattke2022blacksmith}), gain access to confidential data (e.g.,~\cite{kwong2020rambleed, rakin2022deepsteal, jolt23recovering}), or maliciously destroy the safety and accuracy of a system, e.g., an otherwise accurate machine learning inference engine (e.g.,~\cite{hong2019terminal, yao2020deephammer}). The mindset enabled by RowHammer bitflips caused a renewed interest in hardware security research, enticing many researchers to deeply understand hardware's inner workings and find new vulnerabilities. Thus, hardware security issues have become mainstream discussion in top security \& architecture venues, some having sessions entitled RowHammer.

{\em RowHammer Defenses.} Tens of works proposed mitigations against RowHammer, some of which were inspired by the solutions we discussed in our ISCA 2014 paper. To date, the search for more efficient and low-cost RowHammer solutions continues. We refer the reader to our prior overview papers~\cite{mutlu2023fundamentally, mutlu2019retrospective, yaglikci2021blockhammer} and more recent works in 2023 (e.g.,~\cite{marazzi2023rega, wi2023shadow, juffinger2023csi}).

{\em RowHammer Analyses.} Our paper initiated works at both architectural \& circuit/device-levels to better understand RowHammer and reverse-engineer DRAM chips, to develop better models, defenses, and attacks (see~\cite{mutlu2023fundamentally, mutlu2019retrospective}). Our ISCA'20 work~\cite{kim2020revisiting} revisited RowHammer, comprehensively analyzed of 1580 DRAM chips of three different types from at least two generations, showing that RowHammer has gotten much worse with technology scaling \& existing solutions are not effective at future vulnerability levels. 

{\em Industry Reaction: Attacks, Analyses, and Mitigations.} Folks developing industrial memory testing programs immediately included RowHammer tests, e.g., in memtest86~\cite{rh-passmark}, citing our work. Industry needed to immediately protect RowHammer-vulnerable chips already in the field, so almost all system vendors increased refresh rates; a solution we examined in our paper and deemed costly for performance and energy, yet it was the only practical lever that could be used in the field. Apple publicly acknowledged our work in their security release~\cite{AppleRefInc} that announced higher refresh rates to mitigate RowHammer. Intel designed memory controllers that performed probabilistic activations (i.e., pTRR~\cite{frigo2020trrespass,kaczmarski2014thoughts}), similar to our PARA solution~\cite{kim2014flipping}. DRAM vendors modified the DRAM standard to introduce TRR (target row refresh) mechanisms~\cite{frigo2020trrespass} and claimed their new DDR4 chips to be RowHammer-free~\cite{frigo2020trrespass, hassan2021utrr}. This bold claim was later refuted by our TRRespass work~\cite{frigo2020trrespass} in 2020, which introduced the many-sided RowHammer attack to circumvent internal protection mechanisms added to the DRAM chips. Our later work, Uncovering TRR~\cite{hassan2021utrr} showed that one can almost completely reverse-engineer and thus easily bypass RowHammer mitigations employed in all tested DRAM chips, i.e., RowHammer solutions in DRAM chips are broken. The analysis done by our two major works in 2020~\cite{frigo2020trrespass,kim2020revisiting} caused the industry to reorganize the RowHammer task group at JEDEC, which produced two white papers on mitigating RowHammer~\cite{jedec2021nearterm,jedec2021system}. Nine years after our paper, in 2023, two major DRAM vendors, SK Hynix and Samsung, finally wrote papers~\cite{sk-hynix-isscc2023, hong2023dsac} on the RowHammer problem, describing their solutions. Several of these industry solutions build on the probabilistic \& access-counter-based solution approaches our ISCA 2014 paper introduced. 

Major Internet and cloud systems companies also took a deep interest in RowHammer as it can greatly impact their system security, dependability, and availability. Multiple works from Google, e.g., by Google Project Zero in 2015~\cite{google-project-zero, seaborn2015exploiting} and Half Double in 2021-2022~\cite{kogler2022half} directly built on our paper to demonstrate attacks in real systems. Researchers from Microsoft have developed deeper analyses of RowHammer~\cite{cojocar2020rowhammer}, along with new RowHammer attacks~\cite{loughlin2022moesiprime} and defenses (e.g.,~\cite{bennett2021panopticon, loughlin2021stop, saroiu2022configure, loughlin2022moesiprime}).

\section{Summary and Future Outlook}
\vspace{-2pt}

Since 2012-2014, RowHammer vulnerability has become much worse due to technology scaling: without mitigation, one can now induce RowHammer bitflips with orders of magnitude smaller number of activations (e.g., $\sim$10K) and cause much higher rates of errors in cutting-edge DRAM chips~\cite{kim2020revisiting, hassan2021utrr}. Sophisticated attacks are continuously developed to circumvent the mitigations employed in real DRAM chips. Fortunately, we have also come a long way in further understanding and better mitigating the RowHammer vulnerability. The industry is now (hopefully) fully aware of the importance of the problem and of avoiding bitlips. Unfortunately, an efficient and completely-secure solution is not found yet. The solution space poses a rich area of tradeoffs in terms of security, performance, power/energy, cost/complexity. All solutions forego some desirable properties in favor of others. As such, a critical direction for the future is to find solutions superior to what we have today. We believe system-DRAM cooperation~\cite{mutlu2013memory, mutlu2023fundamentally} will be important to enabling complete solutions. We also believe it is critical to deeply understand the properties of RowHammer under many different conditions so that we can develop effective solutions that work under all circumstances. Unfortunately, we do not yet fully  understand many facets of RowHammer (see~\cite{mutlu2023fundamentally, orosa2021deeper, yaglikci2022understanding, luo2023rowpress}). 

DRAM technology scaling will continue to create problems that will exacerbate the bitflips and the resulting robustness (i.e., safety/security/reliability) problems. Our ISCA 2023 paper on RowPress~\cite{luo2023rowpress} provides the first scientific and detailed characterization, analysis, and real-system demonstration of yet another read disturbance mechanism in DRAM. What other fascinating problems will we see and can we completely solve them efficiently? Will we ever be free of bitflips at the system and application levels?

\setstretch{0.78}

\balance
\bibliographystyle{IEEEtran}
{\tiny
\bibliography{combined}}

\end{document}